\documentclass{article}

\usepackage{PRIMEarxiv}
\usepackage{placeins}
\usepackage{dirtytalk}
\usepackage[utf8]{inputenc} 
\usepackage[T1]{fontenc}    
\usepackage{hyperref}       
\usepackage{url}            
\usepackage{booktabs}       
\usepackage{amsfonts}       
\usepackage{nicefrac}       
\usepackage{microtype}      
\usepackage{lipsum}
\usepackage{fancyhdr}       
\usepackage{graphicx}       
\usepackage{adjustbox}
\usepackage{csquotes}
\usepackage{multirow}
\graphicspath{{media/}}     

\title{Not Another Day Zero: Design Hackathons for Community-Based Water Quality Monitoring
}

\author{
  Srishti Gupta \\
  College of Information Science and Technology \\
  Pennsylvania State University \\
   \And
  Chun-Hua Tsai \\
  College of Information Science and Technology \\
  University of Nebraska at Omaha \\
   \AND
   John M. Carroll \\
   College of Information Science and Technology \\
   Pennsylvania State University \\
}

\begin{document}
\maketitle

\begin{abstract}
This study looks at water quality monitoring and management as a new form of community engagement. Through a series of a unique research method called `design hackathons', we engaged with a hyperlocal community of citizens who are actively involved in monitoring and management of their local watershed. These design hackathons sought to understand the motivation, practices, collaboration and experiences of these citizens. Qualitative analysis of data revealed the nature of the complex stakeholder network, workflow practices, initiatives to engage with a larger community, current state of technological infrastructure being used, and innovative design scenarios proposed by the hackathon participants. Based on this comprehensive analysis, we conceptualize water quality monitoring and management as community-based monitoring and management, and water data as community data. Such a conceptualization sheds light on how these practices can help in preempting water crisis by empowering citizens through increased awareness, active participation and informal learning of water data and resources.
\end{abstract}


\section{Introduction}
Water is one of the most critical shared-community resource. Yet, it has been a victim of depletion and deterioration, often a consequence of political and socio-economic turmoil. For instance, Cape Town's `Day Zero' caught international media's attention as a notorious political response to the city's worst drought in over a century \cite{booysen2019temporal}. Unfortunately, research and media explorations into reasons for such crisis, started transpiring after the catastrophe had already happened. Moreover, citizens were not part of the narrative until the very end \cite{le2018cape}. Similarly, the infamous lead water contamination in Flint, Michigan was primarily caused due to the negligence of the local governance \cite{abernethy2018activeremediation}.

These aforementioned examples show that it is imperative for citizens to understand and participate in sustainable conservation practices. Where government and private market policies were considered to be the sole panacea for natural resource management, studies now recognize that it is difficult to formulate effective sustainable policies and practices without involving citizens \cite{agrawal1999enchantment}. Previous studies show that citizens in a community are often self-motivated to participate in such practices because natural resources like water directly affect their well-being \cite{pollock2005community}. For instance, researchers in environmental sciences have framed the practice of citizens, government, academia, and other civic institutions working together to tackle a community problem (often a natural resource), through data collection and analysis, as \textit{community-based monitoring (CBM)}. CBM is hence, a community of practice \cite{wenger2011communities} and is often seen as a part of the citizen science movement \cite{conrad2011review}. Citizen science is a research technique, where researchers involve citizens as volunteers, participants or collaborators, to collect or process data \cite{silvertown2009new}. 

Today there are thousands of community-based water monitoring groups across United States \cite{NWQMC2019}. A recent watershed system study showed how the relatively technical activity of water quality monitoring is sustainable and engaging \cite{carroll2019empowering}. Water quality monitoring occupies a position in nonprofit community work analogous to Habitat for Humanity, Meals on Wheels, volunteer fire companies, time-banks, and food banks. That is, water monitoring is a technical activity that can more broadly strengthen the local community. Watershed systems become measurably stronger with greater citizen-based water quality monitoring activity. In fact, a study of 2,150 US watershed systems by Grant and Langpap \cite{grant2019private} showed that the quality of local water is positively correlated with greater activity of citizen volunteers.

Citizen involvement in such field-based scientific research has the potential to strengthen community by facilitating data-driven informal learning of local environmental problems and resources, increase community attachment and identity, community participation, and engagement \cite{carroll2019empowering}. Emerging research in e-governance and community informatics has also bolstered this belief by advocating for data-driven practices that can be used to facilitate active citizen involvement with the government to jointly tackle environmental issues \cite{linders2012government}. For instance, Carroll et al.  \cite{carroll2018strengthening} initiated the idea of \textit{community data} which is defined as "data pertaining to a community and its locale, that is, data gathered, analyzed, interpreted, and used by members of a local community." Community data aims to empower citizens by emphasizing the need for them to become data-literate so that they can effectively participate in rational decision-making with local government. 

In this paper we describe our work with community-based water monitoring groups of Clear Creek watershed. We identified a hyperlocal community of citizens actively involved in monitoring and management of their local watershed. The community consists of citizen-based groups and municipal authorities who monitor and manage their local water system. This is exhibited as a community of practice (CoP) \cite{wenger2011communities} as it pertains to the domain of water quality, members have built strong relationships with each other, and built formal and informal protocols for water quality monitoring and management. Our goal was to bring this community of practice and other community residents together to brainstorm existing problems in their water system, and design scenarios to address those problems. Therefore, we organized `Design Hackathons' as a participatory design method to create an agile environment for the community to discuss and brainstorm ideas. Design hackathons provided an approach to explore problems and design solutions in an exploratory investigation. We define `Design Hackathons' as time-bounded, issue and design-oriented events, where participants and researchers discuss the issue at hand and explore possible design directions. We conducted seven hackathons with key persons from the water groups identified, and community residents who were not part of these groups. We used three design prompts: mock-up of an integrated water data platform, a stakeholder network diagram, and an interactive learning simulation model, to evoke discussions during these sessions. We explain the rationale and use of these design prompts in detail in `Section 3 Methodology'. The key outcome of these design hackathons was generation of design scenarios by participants. 

Our findings revealed the importance of collaboration among groups in sustaining this community of practice, role of water data, state of current technological infrastructure to manage this data, and role of this community of practice in strengthening community awareness of water system and issues. These insights helped in an iterative development of the mock-up, which transformed from a prompt to a key design outcome of this research. The mock-up transformed into a prototype of a comprehensive networked community-based monitoring platform. This platform had various functionalities like integration of water data across the community, a place for discussions and social interactions for all community members, and learning tools for both adult and young members of the community. Hence, the prototype and design scenarios were the outcomes of our empirical investigation. These design outcomes both complement and augment current research on citizen science in CSCW and HCI. Existing research on citizen science has broadly focused on problems with data quality, data management and understanding volunteer participation in citizen science programs. Hence, our design outcomes compliment the existing research by validating the need of an integrated water data platform for more efficient accessibility, and add to the existing research by producing a more comprehensive platform that incorporates other socio-technical community needs and interactions.

In addition, the analysis of these findings through a community lens demonstrate how this community of practice is not just pivotal in safeguarding local water system but also plays a vital role in strengthening the broader community. Even though design problems and solutions in citizen science have been studied in CSCW and HCI, our study methodically investigated citizen science in the context of a community. We argue that studying such a practice in the context of community illustrates the complex network of actors, their numerous efforts and collaboration practices towards maintaining a critical natural resource - \textit{water}. Design that is driven by such a comprehensive analysis, better incorporates the needs of its potential users, and its elements finely reflect and embody observed and evoked collaborations. Moving beyond the scope of its potential users, such a design further presents opportunities to facilitate increased awareness and knowledge among community members, ultimately leading to active participation by citizens. Water data literacy will allow citizens to think more rationally, make data-driven decisions with respect to water management, and facilitate a rational discourse between citizens and local government. This is especially significant since citizen-government conflict has existed throughout the history of time \cite{coleman1957community}. 

To respect anonymity of the stakeholders, we have substituted names while addressing the water groups and the watershed. For the rest of the paper, we refer to the key people from different water groups as `water stakeholders' and other community participants as `community members'. 

\section{Related Work}
Citizen science is the practice of public participation in science \cite{silvertown2009new}. The citizens science paradigm has become popular in disciplines like ecology, astronomy, and archaeology that typically require large amounts of data for research. Citizens typically participate as volunteers to assist in various scientific activities such as data collection, data analysis, and transcription. Public participation in science not only benefits the research but can also build stronger communities by inculcating informal learning, data literacy, and community engagement \cite{bonney2009citizen}. This can create communities of practice \cite{wenger2011communities} where citizens work together to gather, organize, and analyze critical environmental data, providing a meaningful and accessible path to learning more about data, and putting it to good use. 

\subsection{Citizen science in Design Research}
Citizen science is a relatively new research area in CSCW and HCI research \cite{preece2014hci}. Design research can help improve and support citizen science practices by understanding the socio-technical aspects of this phenomena to build appropriate technical infrastructure. 

For instance, previous research has looked at understanding the motivation for citizen scientists to voluntarily participate in scientific endeavors. Although previous research has established that participation mainly stems from intrinsic motivation, it is also important to understand how and why the nature of participation changes over time \cite{rotman2014does}. Such an understanding will help in the design of better recruitment strategies that will ensure sustained participation \cite{lee2017recruiting, harandi2018talking, jackson2016way}. Secondly, ensuring quality of data collected by citizen scientists requires an efficient data collection and management system. Inefficient data management systems not only leads to poor data quality but also introduce other data challenges such as data interoperability and reuse. In addition, citizen science poses some other pivotal challenges such as communicating findings of a citizen science project to a wider audience, and data visualization, which have now become topics of research interest in HCI \cite{hochachka2012data, mckinley2017citizen, snyder2017vernacular}. Although current citizen science research recognizes these issues, currently no concrete design solutions exist in this space \cite{newman2012future}. Most of these technologies are domain or project specific and design solutions are difficult to appropriate or generalize \cite{soranno2015building}. However, research in this area has produced strong design guidelines to tackle such data problems, \cite{wiggins2013data}. In addition, `democratization' of systems that facilitate citizen science data management, does not require building of sophisticated database and GIS systems \cite{sheppard2012wq}. Instead, research has shown that volunteers are often reluctant to learn new and complex systems. Hence, designers should focus on ensuring interoperability and normalization of data for existing rudimentary software \cite{sheppard2014capturing}. Thirdly, trust in citizen science data can be seen as a byproduct of assurance of usable and good quality data. Hence, research should focus on designing platforms that ensure usefulness of data \cite{kim2011creek, law2017science}.

Besides assisting professional scientists in research, citizen science is a socio-technical activity that strengthens communities by facilitating social capital, informal learning, awareness and increased engagement \cite{conrad2011review}. Hence, citizen science is sometimes referred as `community-based monitoring' where community members work together with various community institutions to protect, conserve and monitor their natural resources. For instance, Burgos et al. \cite{burgos2013systems} presented a conceptual model of community-based monitoring and then applied that model to a particular case of water quality monitoring in rural Mexico. One of the key indicators in this proposed model was social capital, which was empirically and qualitatively evaluated for their case study. Their study found CBM to foster social capital, social learning, and increased participation. Moreover, these community indicators were found to be crucial in taking \say{environmental, territorial or political decisions.}

\subsection{Design Hackathons}

`Hackathons' are traditionally viewed as time-bounded events where participants engage in an ad-hoc and intensive programming spree, and develop a working software or hardware application. A general rationale for hackathons is to create an agile environment for people to generate innovative ideas and working prototypes \cite{frey2016innovation}. Recently, the design community has started to adopt these hackathon-style events to generate design ideas and prototypes  \cite{nolte2018you, trainer2014community, porter2017reappropriating}. 

Most hackathons in the design community are `issue-oriented', that is, these events focus on addressing specific social issues in an ad-hoc manner \cite{lodato2016issue}. For example, Birbeck et al. \cite{birbeck2017self} conducted a hackathon with direct user-engagement to facilitate design innovations for people affected by self-harm. They conducted a hackathon where a team of designers designed technologies for those affected by self-harm. These designs were subsequently critiqued by stakeholders involved with those affected by self-harm.

In the light of following democratic principles advocated by participatory design methodology and \cite{kensing1998participatory} and issue-oriented design, hackathons in the design space have started to engage both technical and non-technical participants. Technical participants here refer to people with expert knowledge on programming languages who would generally be the usual participants for such events, and non-technical participants include non-experts in programming and software/hardware development. Non-technical participants could be domain experts for the issue being addressed but are not necessarily expert programmers. For example, Thomer et al. \cite{thomer2016co} conducted an interface design hackathon to facilitate better software design and support taxonomic work. The hackathon included both technical (programmers) and non-technical (taxonomists) participants.

A natural question here is to understand the role non-technical participants play in rapid design and development activities. Since most design hackathons are issue-oriented and participatory in nature, it becomes important to include stakeholders who would be directly affected by the intended design. Since, non-technical participants cannot directly develop applications, most studies that include such participants provide customized `developmental' kits to participants to design low-fidelity prototypes \cite{taylor2018everybody, taylor2017community, taylor2018strategies}. Despite the popularity of hackathons, design and implementation of such events in the design community remains amorphous. This study attempted to implement issue-oriented and time-bounded design hackathons where participants were members of the local community. The distinctive aspects of these hackathons were (1) it was participatory in nature, i.e., it aimed to empower the local community people who aren't designers or programmers, (2) the term `hackathon' inspired and excited people to participate in designing solutions for their local water quality problems, (3) it led to the creation of design scenarios by participants; and (4) it created a unique synergy by strengthening social ties, facilitating constructive discussions, and exchange of information. 

\section{Methodology}

Our research team had previously worked with Senior Environmental Brigade (SEB), a citizen-based water quality monitoring group comprised of older adults. This gave us a starting point to recruit participants. Subsequently, through a snowball sampling process \cite{noy2008sampling}, we identified thirteen water monitoring groups in the community. We sent emails to key members of these groups, to invite them to participate in hackathon-style participatory design sessions. No financial compensation was offered for participants to join the study. Out of these thirteen groups, six were citizen-based and non-profit groups, and the rest seven were government agencies. 

\vspace{8pt}

\textbf{Citizen-based and non-profit groups included:}

\begin{table}[h]
\begin{tabular}{lll}
1. & Senior Environmental Brigade      & (SEB)   \\
2. & Trout Welfare                     & (TW)    \\
3. & The Atlas Group                   & (Atlas) \\
3. & Water Conservancy Collective      & (WCC)   \\
5. & Clear Creek Watershed Association & (CCWA)  \\
6. & Resource Monitoring Roundtable    & (RMR)  
\end{tabular}
\end{table}
\FloatBarrier

\textbf{Government agencies included:}

\begin{table}[h]
\begin{tabular}{lll}
1. & Beneficial Reuse Authority            & (BRA)  \\
2. & Beauville Town Water Authority        & (BTWA) \\
3. & Beauville Facilities Office           & (BFO)  \\
3. & Clear Creek Watershed Commission      & (CCWC) \\
5. & State Environmental Protection Agency & (SEPA) \\
6. & United States Geological Survey       & (USGS)   \\
7. & Regional Municipalities               & (Municipality) 
\end{tabular}
\end{table}

\subsection{Design Hackathons Procedure}

Design hackathons were designed as participatory events, that aimed to empower local community residents, who aren't professional designers or programmers. Since this is an exploratory research, design hackathons provided an inductive approach in investigating the problems with the community's local water system, and at the same time brainstorm design solutions with both the water stakeholders and community residents. The hackathons were participatory as they provided a space to all participants to voice their opinions and ideas, and contribute to the design process. The events were agile, and the term 'hackathon' also persuaded and excited people to participate in the design thinking process.

Unlike contemporary studies on hackathons, where the focus is for participants to create a tangible artefact or prototype, our focus was to engage participants as equal stakeholders in the design process, and give them the agency to participate in design thinking. Various discussions and brainstorming sessions lead to the creation of \textbf{\textit{scenarios for design}}, which was the primary outcome of the hackathons. This study draws from, and argues for research through design paradigm, where design is not just creation of an artefact, but incorporates the entire process and activities to provide explanations and solutions to a problem \cite{haynes2009design}.

We conducted seven design hackathons. Each hackathon lasted for an average of two hours. Out of these seven hackathons, four sessions were conducted with the water stakeholders, i.e., key persons from water monitoring groups, and the rest of the three sessions were open to all members of the community to participate. All hackathon participants were non-technical and included either domain experts (water stakeholders) or community residents. Table \ref{tab1} provides an overview of the hackathons in terms of the participants, focus of each hackathon session, and outcomes generated from each session. 

\begin{table}[h]
\centering
\begin{tabular}{|c|l|l|l|}
\hline
\textbf{Hackathon} & \multicolumn{1}{c|}{\textbf{No. of Participants \& Groups}} & \multicolumn{1}{c|}{\textbf{Session Focus}} & \multicolumn{1}{c|}{\textbf{Outcome}} \\ \hline
H1 & \begin{tabular}[c]{@{}l@{}}7 Participants - \\ Water Stakeholder\\ (RMR, WCC, TW, \\ SEB, Municipality)\end{tabular} & \begin{tabular}[c]{@{}l@{}}- Introductions\\ - Problems explored\end{tabular} & \begin{tabular}[c]{@{}l@{}}- Stakeholder Network\\ - Water data platform mock-up\end{tabular} \\ \hline
H2 & \begin{tabular}[c]{@{}l@{}}6 Participants - \\ Water Stakeholders\\ (WCC, SEB, \\ CCWA, Atlas, BRA)\end{tabular} & \begin{tabular}[c]{@{}l@{}}- Understanding stakeholder network\\ - Exploring mock-up \\ for design solutions\end{tabular} & \begin{tabular}[c]{@{}l@{}}- Improved stakeholder \\ network diagram\\ - Proposed scenarios for design\end{tabular} \\ \hline
H3 & \begin{tabular}[c]{@{}l@{}}5 Participants - \\ Water Stakeholders\\ (BRA, SPWC, Municipality, \\ Public Library)\end{tabular} & \begin{tabular}[c]{@{}l@{}}- Presented improved \\ stakeholder network\\ - Presented improved mock-up\end{tabular} & \begin{tabular}[c]{@{}l@{}}- Problem scenarios for design\\ - Improved mock-up\end{tabular} \\ \hline
H4 & \begin{tabular}[c]{@{}l@{}}18 Participants - \\ Community Residents\end{tabular} & - Presented improved mock-up & - Problem scenarios for design \\ \hline
H5 & \begin{tabular}[c]{@{}l@{}}4 Participants - \\ Community Residents\end{tabular} & - Presented improved mock-up & - Problem scenarios for design \\ \hline
H6 & \begin{tabular}[c]{@{}l@{}}5 Participants - \\ Water Stakeholders\\ (BUFO, Atlas, CCWA, SEB)\end{tabular} & - Evaluation of scenarios & \begin{tabular}[c]{@{}l@{}}- Requirement analysis \\ of scenarios\\ - Proposed design of \\ interactive learning \\ simulations for scenarios\end{tabular} \\ \hline
H7 & \begin{tabular}[c]{@{}l@{}}5 Participants - \\ Community Residents\end{tabular} & \begin{tabular}[c]{@{}l@{}}- Presented an interactive\\  simulation model\end{tabular} & \begin{tabular}[c]{@{}l@{}}- Feedback on interactive simulation \\ - Proposed scenarios for design\end{tabular} \\ \hline
\end{tabular}
\vspace{2pt}
\caption{Overview of Hackathons}
\label{tab1}
\end{table}

The seven hackathon sessions can be viewed as a vector of change in development from less specified ideas to more concrete design possibilities. The sessions were designed progressively, i.e., the course of change shifted from being relatively broad focus on identifying issues, directions and possibilities (H1), to more focused ideas (H2) and requirements (H3, H4, H5), and finally more specific walk-through of design (H6, H7). 

The first hackathon (H1) was an exploratory discussion session where we first introduced the objective of the project as an exploratory study to understand how water data investigations could help citizens manage and improve their community's local water resources. The water stakeholders introduced themselves, explained the mission of their organizations, current workflow of data practices, and issues they face with respect to water data and engaging with the community. The discussion in first hackathon, gave us two important insights about this community of practice: 
\begin{enumerate}
    \item Water groups collect different kinds of data, and share this data with each other for various purposes.
    \item There are different kinds of collaborations among water groups to support this community of practice.
\end{enumerate}

\begin{figure}[h]
  \centering
  \includegraphics[width=\linewidth]{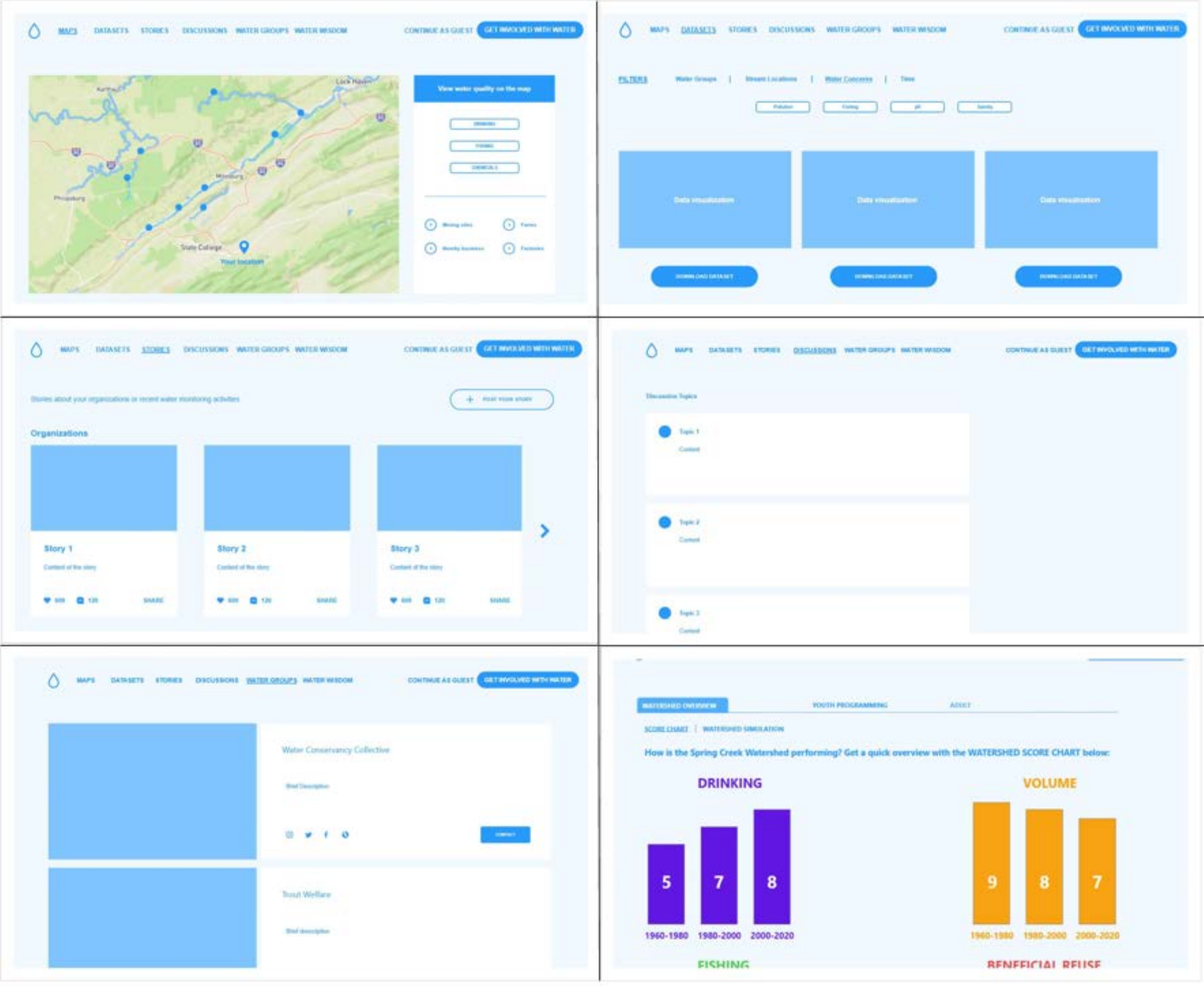}
  \caption{Water data platform mock-up} 
  \label{figure:mockups}
\end{figure}

\begin{figure}[h]
\centering
\includegraphics[height=2in, width=4in]{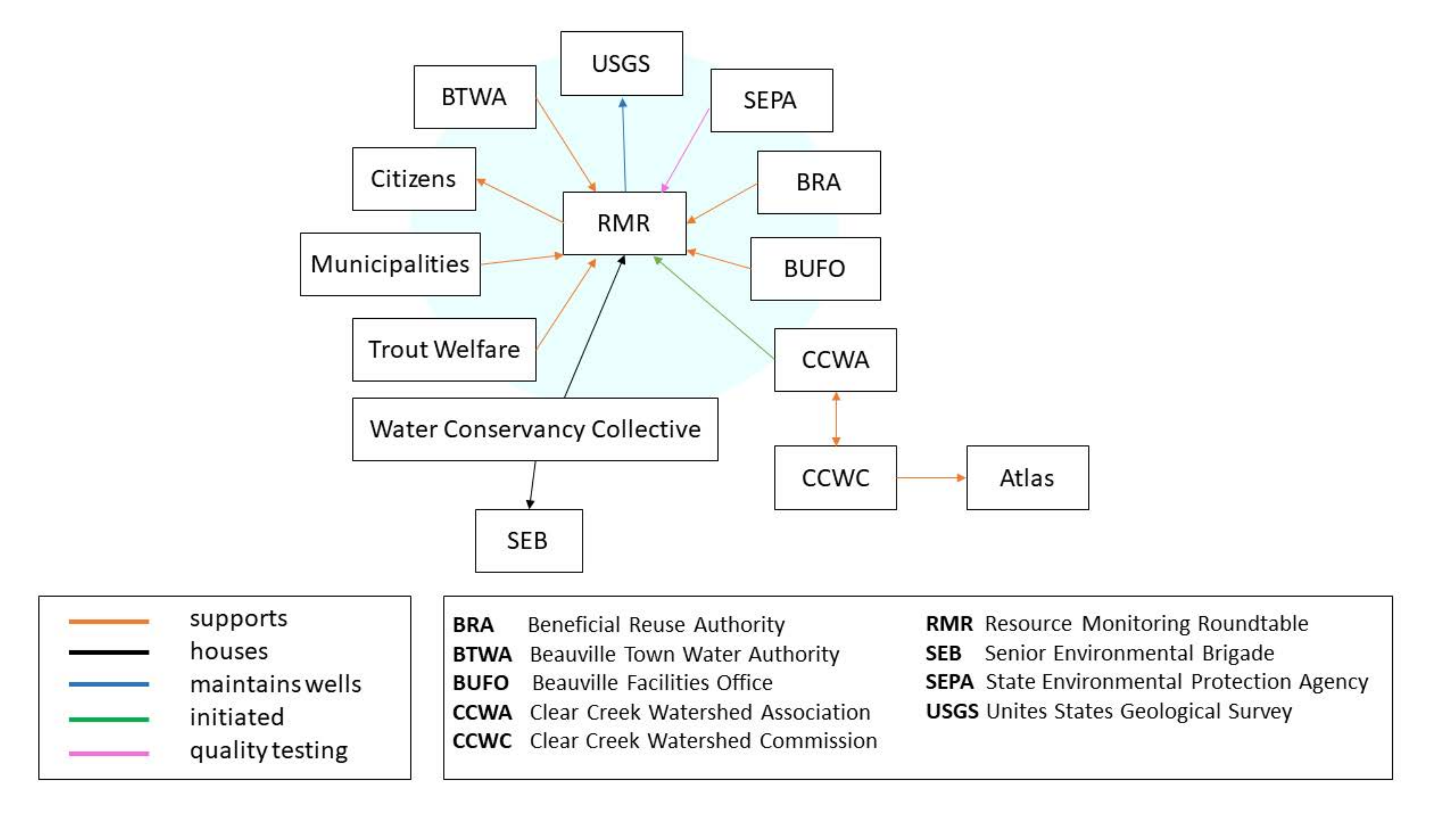}
\caption{Stakeholder Network} \label{figure:network}
\end{figure}

These insights gave us a direction to organize the subsequent hackathons. We attempted to illustrate the two insights through design and created:
\begin{enumerate}
    \item a mock-up of a water data platform to represent data collected by the water groups (Figure \ref{figure:mockups})
    \item a stakeholder network diagram (see Figure \ref{figure:network} to better identify all water stakeholders and the relationship between them.
\end{enumerate}

These two designs were then used as prompts to evoke discussions in the subsequent hackathons. 

The initial design of the mock-up had a map-based interface for a database that integrated water data collected by different groups(top-left screen in Figure \ref{figure:mockups}). The design eventually evolved based on participant's feedback and discussions. We subsequently added an easy filter model to access this integrated data (top-right screen in Figure \ref{figure:mockups}), a space for citizens to post stories about events or their experiences in the Clear Creek Watershed (middle-right screen in Figure \ref{figure:mockups}), a discussion forum where citizens could ask questions to the water stakeholders (middle-left screen in Figure \ref{figure:mockups}), a contact and information page of all local water groups in the community (bottom-right screen in Figure \ref{figure:mockups}), and an informal learning page that fosters data-driven learning about water in the local community (bottom-left screen in Figure \ref{figure:mockups}).
Even though the mock-up was used as a prompt to evoke discussion, we emphasized that it was not to be misunderstood to be a fixed prototype or design idea. We encouraged participants to think and express ideas beyond the prompt as well.

The stakeholder network diagram (see Figure \ref{figure:network}) was also presented in the hackathons to get feedback from water stakeholders, and validate our understanding and representation of connections among different groups. Incidentally, this prompt turned out to be an important cue as it not only helped us understand different types of collaborations among groups, but also led to proposition of new scenarios, and new ideas to improve the mock-up.

From the hackathons 2 to 5, we continued to iteratively improve the mock-up and stakeholder network and evoke discussions through these two prompts. These four hackathons led to generation of a number of scenarios for design, which were analysed and collated to five concrete scenarios. Together, these were evaluated in hackathon 6 with water stakeholders. The discussion and evaluation led to the conclusion that design of these scenarios could serve as an eminent evidence-based and informal learning tool for the community, to learn about various water issues in their locale. The evaluation also subsequently helped deconstruct and elicit requirements for design of each scenario. Water stakeholders and our research team eventually decided that these scenarios should be designed as interactive data-driven simulation models, to effectively educate and spread awareness in the community. 

\begin{figure}[h]
  \centering
  \includegraphics[width=\linewidth]{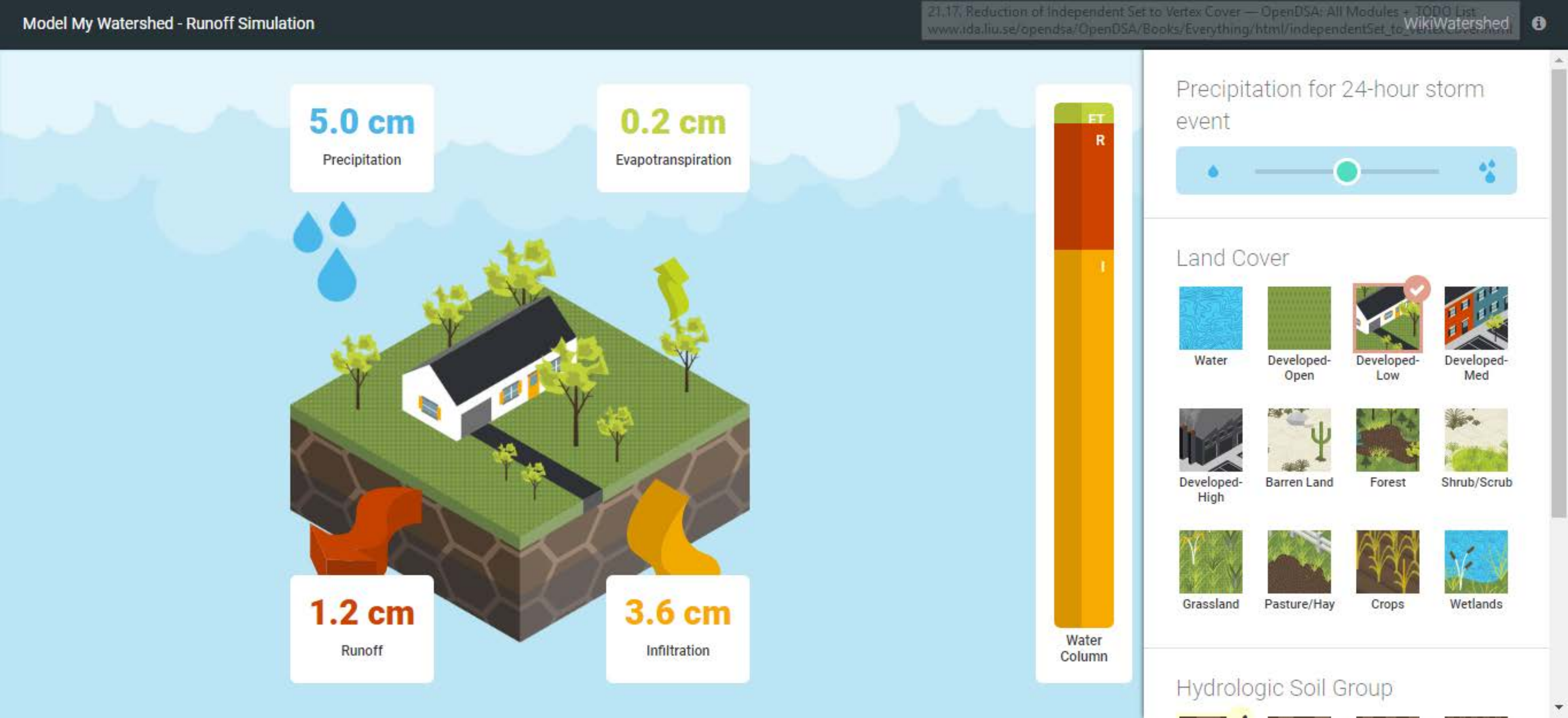}
  \caption{Runoff Simulation Model Presented in Hackathon 7, via Stroud Research Center. (Source: \url{https://runoff.modelmywatershed.org/})} \label{figure:runoff}
\end{figure}

However, before building these interactive simulation models, we decided to test our design idea with our community resident participants to understand whether such a design would be an effective learning tool or not. For this purpose, we searched for interactive simulation tools for water quality, and discovered an interactive runoff simulation model developed by the Stroud Research Center \footnote{https://stroudcenter.org/}. This model was a simple animated simulation that taught users how land use and soil affected rainfall in a community. We decided to present this model in our final hackathon to discern participants' reaction towards such a design (see Figure \ref{figure:runoff}). This helped us understand the feasibility of the design idea and also gave community resident participants another prompt to generate new design scenarios. 

\subsection{Data Analysis}
This study used a thematic analysis method to uncover themes from the qualitative data collected during Design Hackathons. Thematic analysis is a qualitative data analysis method that is used to identify themes across a dataset \cite{braun2012thematic}. Thematic analysis can be used for both deductive analysis and inductive analysis. The flexibility of this qualitative data analysis method and its ability to organize and identify meaningful themes across a dataset, were two main reasons to use this method in our analysis. 

This research was exploratory and required inductive analysis of data to identify themes around community-based monitoring in Clear Creek Watershed. The design hackathons were recorded with the consent of the participants, fully transcribed and analyzed by the first two authors. The data was first qualitatively coded using open coding to identify broad categories. Since grounded theory is an iterative approach, the two authors went back and forth between categories and transcripts to identify more specific categories and sub-categories. These identified categories were then deliberated with other authors to finalize the main themes. These themes were organized using axial coding, which is a qualitative coding method to organize identified categories and sub-categories to create a comprehensible model of the data \cite{creswell2007qualitative}. These final themes are organized and elaborated in the Results section below.

\section{RESULTS}
Data analysis revealed six key findings about the nature of different water groups, their practices, and the scenarios for design generated by participants. The nature of different water groups is explained in section 4.1 which delves into understanding the nature of the watershed stakeholder network. This in-depth analysis helps identify and understand the direct stakeholders of future design. The practices of these stakeholders are explained in the following sections. Section 4.2, 4.3 and 4.4 describe the data practices, role of technology in their practices, and initiative taken by these groups to engage with the broader community. Section 4.5 discusses how their role and stakes in the community as figure-heads of the watershed affects their interaction with the community. Finally, section 4.6 discusses the scenarios generated and evaluated by participants.

\subsection{Understanding the Watershed Stakeholder Network}

The Clear Creek Watershed is a region rich in water resources and active and concerned citizens. The stakeholders identified are key persons belonging to different water quality monitoring groups. They helped us learn about the unique functions, connections, and collaboration practices among different groups. As explained in section 3, we developed a stakeholder network diagram (Figure \ref{figure:network}) to elucidate these functions, connections and collaborations among different groups. This stakeholder network diagram was iteratively improved after getting feedback from hackathon participants.
Understanding this stakeholder network is important because it has direct implications for design of water platform. It helps the researchers unfold the `stakes' of the direct users of the intended design. Moreover, it directly embodies the ethical responsibilities advocated by the participatory design methodology, that is, the inclusion of users in the design process.

\subsubsection{Collaboration Among Groups}

The Clear Creek Watershed is a hyperlocal community comprised of various residents, non-profit, and government groups. The water groups in this hyperlocal region have close but complex collaboration practices. The stakeholder network diagram helped in aiding the hackathon discussion in unravelling the nature of these collaborations. The collaboration between water groups can be divided into two broad categories - formal and informal. Formal collaboration includes data sharing and financial support. For instance, Beauville Fish Authority pays USGS to help in data collection that it uses to assess water quality to support fish in the watershed. Informal collaboration on the other hand includes unofficial assistance that watershed stakeholders offer to each other without any conditions. For example, some stakeholders volunteer to help out other groups just as a form of community service or because they share close social relationships with one another. 

\say{\textit{And there are contributors that don't generate data. Like XYZ Township gives money to RMR simply as a feel good effort. Because you recognize that you will get something good out of it someday.}} (Water Stakeholder, H3).

This informal collaboration between the township and RMR exemplifies the close-knit relationship between groups, where one group might provide financial assistance to another group because of their strong social ties and not a formal obligation.

Other kind of collaboration includes government organizations working to help the local citizen and non-profit groups. This kind of collaboration is neither formal nor informal. For instance, the Conservation District serves as a liaison to many different groups in the stakeholder network. Similarly, SEPA has a historical relationship with RMR, therefore, it pays to conduct analysis of data collected by RMR. This is beneficial for the entire community because SEPA has access to very high quality labs. Hence, the watershed stakeholders benefit from access to this high quality water quality data and analysis.

\subsubsection{Functions of Different Groups}
The stakeholder network diagram was also helpful in fostering discussion on better understanding roles and functions of different water groups in the Clear Creek Watershed. For instance, RMR is depicted in the center of the stakeholder network because it serves as a central repository of baseline water quality data that many other water groups in this community use. The discussion also helped understand some interdependent functions of groups, like that of Clear Creek Watershed Commission (CCWC) and Clear Creek Watershed Association (CCWA). CCWC is a group made up elected officials from neighboring fourteen municipalities. The aim of this group is to come up with new ideas, and make decisions about new programs to improve the water quality in this watershed. On the other hand, CCWA is a citizen group made up of volunteers who assist the Clear Creek Watershed Commission by helping them shape and work through new program ideas. They also generate new projects and programs ideas, which, however required approval from the CCWC.

\subsubsection{Improving Stakeholder Network}
The initial design of the stakeholder network diagram was developed based on our interactions with the watershed stakeholders in the first hackathon. The design hackathons helped in getting feedback on this initial design and led to an improved stakeholder network diagram (Figure \ref{figure:network}). The feedback received helped in improving the diagram in two ways. First, it corrected the structure of the diagram and connections between water groups. Second, it gave new ideas on how to better depict the complex nature of collaborations between different groups. For instance, there are a lot of regulatory and municipal groups that needed to be included in the scope of Clear Creek Watershed. Including new stakeholders would require significant restructuring of the stakeholder network. Participants gave suggestions on different ways to better represent the complex stakeholder network. 

\say{\textit{We could include in these boxes, potentially more information about, or with different colors or something, like, who's actually collecting data.}} (Water Stakeholder, H2).

For instance, one water stakeholder suggested using different colors and including more information in the stakeholder network diagram about each group. Hence, participants' input was helpful in both correcting and improving the design of the stakeholder network.

\subsection{Multifaceted Nature of Data Collected by Water Groups}

The data collected by water groups is diverse and fragmented. This is because of the diverse nature and function of each group. This makes the data multifaceted as it affects the way data is used, shared and integrated among groups. We delve into each of these characteristics in this section.

\subsubsection{How is water data used by different groups?}
The Senior Environmental Brigade (SEB) is a citizen-group that primarily collects macroinvertebrates data and posts it openly on their website. The group has a quality control and assurance team that verifies data collected by each team within the group, and a training team that trains new members on data collection practices. Despite following a legitimate procedure, their data cannot be used by government organizations because the group members are not trained scientists and the data collection and analysis is not done in a formal certified environment. On the other hand, data collected by government organizations like BRA, is used for their internal analysis and decision making, and also sent to SEPA for regulatory compliance. Some government groups often involve residents as volunteers to help in data collection. For example, USGS partners with non-profit groups like Trout Welfare to recruit volunteers as citizen scientists. Municipalities also produce water quality data through the Municipal Stormwater Program.  Another unique way data is used can be seen by the example of Resource Monitoring Roundtable (RMR). RMR is a committee formed by representatives of many government, non-profit and municipal organizations, that collect baseline water quality data which is used by various groups in the stakeholder network. Similarly, Municipal Stormwater Program does not directly collect data but its activities are centered around data.

\say{\textit{Occasionally we go and do actual macroinvertebrate study...every now and then we say this year is representative of a dryer year, so let's go do the sampling now to see what it looks like...and then hopefully the next time we have another stream year, that looks very similar to that, we can go take that same macroinvertebrate data and compare the two, and say what has happened in that 5 year-7 year period of similar water years.}} (Water Stakeholder, H1). 

The above quote is an example of how one of the ways RMR baseline data can be used to compare and predict trends in water quality over time.

\subsubsection{How is water data shared between different groups?}
The way data is collected and then shared between different groups is not a linear process. Hence, even though most of the data collected is openly available for the citizens to view and access, the non-linear flow of data between groups makes data accessibility difficult. For instance, most government organizations have to maintain and send data to SEPA for regulatory compliance. This regulatory compliance data is primarily obtained from RMR. Hence, citizens can get the same data from different venues - request this data from RMR, the government organization or access it at the SEPA website. 

\subsubsection{Can we integrate this vast and fragmented water data?}
Different water groups, whether citizen, non-profit or government group have different vision and goal of collecting and using data. Hence, there exists enormous amounts of data in varying formats and meta-data, that would be arduous to integrate in a streamlined format. However, at the same time, participants recognized the need for an integrated database, to expedite their day-to-day data exchange and sharing practices. For instance, a popular suggestion was to start by automating RMR's data, as it is the central repository of baseline water quality, and then slowly integrate it with other widely used data sources like SEPA and USGS:

\say{\textit{If the goal is to try and get all of the water quality and quantity data, easily accessible by citizens and anyone else to look at it, my suggestion would be to just pick a few of those and get it to work, and then add in the other ones, because it's just a monstrous amount of data and not everybody keeps their data the same way.}} (Water Stakeholder, H3).

In addition, there exists some data redundancy, as different groups tend to collect data in the same streams, but at slightly different locations on that stream. Current infrastructure does not have the capability to incorporate such redundancies in a meaningful way. If an integrated data platform is designed to allow such a meaningful incorporation, it will enable analysts to use this redundant data to verify any kind of discrepancy. 

There have been some public initiatives to integrate data into a single platform. For example, the Hydraulic Network is a research group in Beauville Township that integrates data collected by SEPA, USGS and it's partner group Alliance for Water Resource Monitoring (AWRM), (research group initiative in a college near Beauville University) in a separate database platform. Their data integration is focused on bringing together data that can help identify water resources that have been affected by gas exploitation. This is one of the very few initiatives to integrate massive amounts water data, but it is limited in its focus.

\subsection{Technological Infrastructure to Supports Current Practices}
The hackathon discussions helped us understand the current state of affairs with respect to technology use in the watershed. The participants talked about the state of their current technological infrastructure and technology initiatives of organizations outside of the watershed. Since one of the prompts of the mock-up was the idea of citizen learning about water quality, the participants discussed with us and each other about different education technology. For instance, an independent water research group, Stroud Research Center, outside of this watershed, developed an application called ModelMyWatershed (see Figure \ref{figure:runoff}) to engage citizens in learning about how a watershed works. Similarly, one participant talked about their university's drinking water quality app, which can be used by all university staff, faculty and students to monitor drinking water quality within the university facilities. The participants also reached a consensus and recognized the importance of a water platform for educating school children. 

\vspace{2.0pt}
Currently, water groups use hardware and software applications to manage their monitoring activities. Hardware technology includes monitoring devices that are used to collect water quality data, and software technology includes database applications and quality assurance \& control (QA/QC) applications. The discussions revealed various problems associated with this current infrastructure. 

Firstly, data collected requires to go through quality control and assurance (QA/QC) check before it can be shared and analyzed. Monitoring devices work in a very narrow band. Current protocols for QA/QC requires the person in-charge to manually go through the dataset and correct any offset value or outliers. There is no automated mechanism to clean the data. The following quote provides an example of such a problem. 

\say{\textit{Let me give you a picture of what happens when you want data from the RMR. So you call the Water Conservancy Collective's office and you ask for data or you send them an email and say `I want this data'. And if you're not asking for a lot of data you might get it the next day. Usually I am asking for `Give me the data for last 16 years of this, or these 4 stations', and so then they have got to go through and figure out how to get the data out of the database, check and make sure it's all correct and finally give it to me. Well that can take months. If you're a citizen looking for data, you don't want to wait a month!}} (Water Stakeholder, H3).

Even though RMR maintains the central repository of baseline water quality data, it does not have a streamlined QA/QC system. When a requester requests this data, RMR starts the QA/QC or the `cleaning' process. This process can take any amount of time, depending on the size of the data:

Secondly, most groups either use rudimentary spreadsheets to store their data, or use databases that are not compatible with other database technologies making it difficult to share data. For instance, one participant explained the problem of interoperability caused by the use of basic spreadsheets to store data. The participant explained that when BRA has to send data to SEPA, lack of interoperability leads SEPA to manually transfer data between the two systems. 

\say{\textit{We put it (data) in an excel spreadsheet, and then we would send it to SEPA, and they would then take our numbers off of our spreadsheet and enter it into their system by hand.}} (Water Stakeholder, H3).

These problems reveal the lack of appropriate cyber-infrastructure to support day-to-day activities of water groups. It is astonishing how a hyperlocal community of citizens have created a somewhat loosely coupled socio-technical system. The vast data collected by these groups has enormous potential to drive change, but lacks the appropriate technological infrastructure to foster their practices.

\subsection{Initiatives by Water Groups to Engage with the Broader Community}
The water groups as a part of the Clear Creek Watershed community, have made several efforts to engage with the residents by creating volunteer programs, inviting residents to participate in events organized by water groups, and engage and involve school and university students in various monitoring and educational programs. Other forms of citizen initiative include jointly working together as a community to create measures to improve efforts of various groups, and efficiency of water monitoring and management projects. The various initiatives can be classified as initiatives taken by citizen/volunteer groups, government groups, and joint initiatives.

\subsubsection{Initiatives by Citizen/Volunteer Groups}
The Clear Creek Watershed Atlas is a citizen initiative that aims at communicating the story of Clear Creek Watershed to the citizens. Their initiative started as a coffee-table book idea, and later transformed into a website with a number of articles written by water stakeholders of this community. The Atlas is the sole resource on the internet that explains the story of the Clear Creek Watershed from a holistic perspective. The Atlas group has vision to transform their website from just static articles to an interactive data and map-based platform to better disseminate information and engage residents. However, they lack resources, knowledge and infrastructure to materialize their ideas. The following quote shows the current ways in which the Atlas group is working towards engaging with the local community and at the same time improving their website.

\say{\textit{..students [from agricultural sciences college] did story maps for the Clear Creek Watershed Atlas. Basically, several of them used some of the data from the Resource Monitoring Roundtable (RMR), and they were able to characterize stream quality in different locations using this data.}} (Water Stakeholder, H2).

Engaging with local university students to create story-maps helped in improving their website, and also strengthened ties with the local community.  

\subsubsection{Initiatives by Government Groups}
Government initiatives are more varied in nature, as they include enforcing policies that will incentivize residents to become proactive with water conservation and management, and develop programs to more directly engage residents in their work. For instance, the municipalities designed a policy for stormwater assessment in neighborhoods. The idea behind this policy is that, if the assessment of a particular residence is found to be unsatisfactory, a tax will be levied. Similarly, the Municipal Stormwater Program involves resident volunteers in projects such as construction of riparian buffers along a stream to prevent sediment flow into the water. On the other hand, every year BRA hosts fifth grade students from local schools in a two week environmental tour. The students get a chance to tour the facilities, look at the data collected and also analyze this data. The goal is to get students excited about water resources and inspire them to look at these organizations as valuable future workplaces.

\subsubsection{Joint Initiatives by Citizen and Government Groups}
In this hyperlocal community, both citizen/volunteer groups and government groups often work together on various projects in order to utilize each other's capabilities and resources. The Resource Monitoring Roundtable (RMR) is one successful example of such joint initiative. The watershed community recognized the need for a good quality central data repository that everyone in the community could use. The RMR is thus a committee formed with representatives from various water groups in the Clear Creek Watershed. These representatives contribute funds to keep the committee running. The committee basically collects baseline water quality data which can be used to analyze the health of the entire watershed, in addition to the individual use of each group. The following quote describes RMR as a joint community initiative:

\say{\textit{...real pulse of the watershed...the idea behind RMR was always that this is a central repository for real data, that we the contributors all agree is good data.}} (Water Stakeholder, H3).

Similarly, the water groups recently started an integrated water management plan. The aim of this plan is to integrate the resources and capabilities of every group in a meaningful way to be able to address different water-related problems together as a team. Example of such problems include, preparing a water budget, dealing with financial cost of various environmental issues, and dealing with political issues and community conflicts.

\subsection{Impediments in Public Discussion}

The water data platform mock-up presented to the participants, had two features separate from data, called `Stories' and `Discussion' (Figure \ref{figure:mockups}). The intent of `Stories' was for groups to post accounts of activities, events, or any other significant instance, that would help create awareness about water groups, citizen engagement initiatives, inspire people to engage with their local water resources, and learn about water quality. The `Discussion' section of the platform was intended to create a space for residents and groups to engage in discussion about questions or concerns residents might have about water. 

Through our discussions with participants, we learned about various obstacles in creating such online public spaces. For instance, the `Discussion' feature was regarded to be a controversial space for water groups since it would require them to take a stand on certain issues in public discussions. It would also require continuous monitoring to ensure people don't post unnecessary and negative content.

\say{\textit{...somebody puts in there something about Nestle water...SEB is gonna respond to that? Or TW? on a blog? I mean my point is that that's what that is. I can put any topic in that I have a question about..first of all you don't know who the questioner is...number two, (volunteer groups) worry about volunteers...and we also have a constituency that funds this. And when issues come up you have problems with funders...}} (Community Resident, H5).

Similarly, for `Stories' it was important to know the provenance of the content and who was going to write these stories. Both citizen and government water groups see themselves as front runners of water management and monitoring in the community, and did not want to take a position on controversial issues. 

\say{\textit{There are questions about how much we want to highlight controversial issues...if we have a story and it looks we wrote it, it's almost like we're taking a position on it...another way to do is to link to stories that are current issues or something. We will be just linking to information and not taking a stand.}} (Water Stakeholder, H2).

These insights eventually led to the removal of the `Discussion' feature, and `Stories' was modified to include only community resident stories.

\subsection{Scenarios for Design Proposed by Participants}
An important goal of these hackathons was to `hack' design ideas collectively with the participants. Even though, the participants were not designers, the discussions prompted by the mock-up and stakeholder network diagram let to an organic generation of concrete scenarios for design. We collated these scenario-based design ideas and validated them with water stakeholders in later hackathons.  

\subsubsection{Scenarios Generated by Participants}
The mock-up of the water data platform prompted participants to think of all the different possibilities of such platform beyond just data integration. For instance, one scenario was about designing affordances to engage school kids more with the watershed community. The idea was to allow school kids to upload water data collected by them on field trips and then be able to view that data on a map on the web platform. If school children could see value of the data they collected in a tangible form on a web platform, it would motivate them to learn and engage more with the watershed community. Another emergent idea by water stakeholders to engage residents was to create a visual 'report card' showing overall health of the watershed. It would be a very simple and easy to understand concept for residents to look at it and make sense of current state of the watershed.

\say{\textit{I could look at it (report card) and say, 'Oh it's got an A in this and B in that'. And you know, you can look at it very quickly and say `Oh it's really good here, but boy it's got a D there ! Maybe we ought to look closer at that and link to data or some way of extracting analysis of data or something.'}} (Water Stakeholder, H2).

In addition, participants posed intriguing questions which serve as important design directions. For instance, if we want to make the data publicly available to the residents, we need to clarify what kind of data we want residents to have access to? What kind of data is already publicly available? How would residents make sense of this data? Can the platform be designed to allow residents to analyse the data? If the aim is to enable resident learning about water data and analysis, the water data platform should provide guidance on how to analyze data and draw insights from this analysis. 

On the contrary, hackathon participants who were not water stakeholders had more specific concerns which they wanted represented in the water data platform. One participant had a scenario where someone would try to figure out the causal relationships to figure out what is going on. He emphasized indirect and systemic effects - differentiating complex causality from simpler relationships like water quality problems occurring near mines. Some participants were interested in using the water platform to reach out to the right water stakeholder to help answer their questions. For example, one participant said that she was concerned about activities of a mining company in the creek near her house. She would want to use such a platform to reach out to the water groups who collect data at that particular sight and understand the impact of mining activities on water quality.

\subsubsection{Learning and Validating Scenarios}
We analyzed the discussions from the hackathon 2,3,4, and 5, and grouped them into five scenarios for design. These scenarios were validated with water stakeholders in hackathon 6.
\paragraph{\textbf{Scenario 1: Planning Riparian Buffer Sites}}
Certain trees are planted near the streams to provide shade and control water temperatures. However, these buffers cannot be assigned randomly. Using temperature and GIS data, you will determine the optimal location for planting. 
\paragraph{\textbf{Scenario 2: Understanding Chemical Impact}}
Beauville Department of Transportation spraying chemical on a property near a Branch of Clear Creek: possible use of water data platform to get data on that particular site to understand how the chemicals may affect water quality.
\paragraph{\textbf{Scenario 3: Measuring Thermal Pollution}}
Thermal pollution can be a problem in surface water, and it endangers trout and other stream life. Using water temperature data, you will interpret temperature, trends, identify possible locations where thermal pollution is happening, and propose solutions.
\paragraph{\textbf{Scenario 4: Assessing Land Development Impact on Water Quality}}
You are concerned about changes to nearby roads and the development of adjacent farms to residential areas as they relate to long-term viability of a stream site. Using the data platform you will track key water quality metrics and explore data-driven questions about possible changes to water quality if specific kinds of development happen in specific locations.
\paragraph{\textbf{Scenario 5: Beneficial Reuse}}
Beneficial reuse water is wastewater that has been processed to meet state and federal regulations for cleanliness, but many community members can't get past the fact that their kids might play soccer on a field that has been irrigated with \say{toilet water}. Using the data platform, you will learn about beneficial reuse, and analyze data describing beneficial reuse water before and after treatment.
\newline

The water stakeholders assessed the validity of the design scenarios, and decided to represent these scenarios through a data-driven learning design for the community. Therefore, it was decided to design interactive learning simulation models to educate the community on the water quality scenarios identified. With this design direction in mind, we aimed at having a comprehensive discussion to understand each scenario in detail. For instance, for the first scenario, it was emphasized that it was important to give background information and rationale about why and how temperature control is important in buffer count plantation. Other details like ownership of property was an important detail to consider in planning optimal location for planning of riparian buffers. Moreover, for the second and fourth scenario, it was important to design the model in a way that residents understand impact of urban development and spraying of chemicals in a constructive way. Not all chemicals and urbanization is bad for the environment and water quality. Residents generally equate urbanization and development with environmental deterioration, and advocate for less development. However, if growth plans are made with best management practices, it can help reduce the negative impact of development. 

\say{\textit{So the challenge here is that there are so many other factors that aren't in this database. For example, how many houses are connected to the wastewater treatment system. You may actually see an inverse relationship of what you may think it should be. We see that water quality decreases as we go back in time.}} (Water Stakeholder, H6).

The above example given by water stakeholder elucidated the importance of having a data-driven model to educate the community and clarify misconceptions that the community might have about their water system. This will in-turn facilitate more rational discourse and citizen engagement in the community.

\subsubsection{Exploring Feasibility of Interactive Learning Models}
Before diving into development of scenarios identified, it was important to investigate whether such interactive learning simulation models would be helpful for the community to learn about water quality, and if users could perceive long-term adoption of such a system. Hence, our next goal was to find a water quality learning simulation model, that could be appropriated and presented to the community for feedback on such a design. Incidentally, in our second hackathon, one of the water stakeholders had given an example of a Runoff Simulation Model developed by the Stroud Research Center (see Figure \ref{figure:runoff}). The model perfectly depicted the design we had envisioned for our scenarios. The interactive model was a simple animation that allows users to learn the effect of amount of rainfall, type of land cover, and type of soil on rain-water runoff and infiltration. We appropriated this model by creating scenarios in the context of Clear Creek Watershed to evoke participants to think about ways to navigate problems represented in the scenarios. For instance, the following scenario presented, represents the problem of urban development, and how a Runoff Simulation Model could be possibly helpful in understanding the impact of such development on groundwater levels.

\begin{displayquote}
\textit{Recently, Beauville Township has seen a considerable amount of urban-development. This increasing development has concerned residents that it might deteriorate land and water resources. It is important to understand how these development practices are carried out and what kind of best management practices are being used by the authorities. According to some water resources experts, water quality in the Clear Creek Watershed has actually improved over the years and is better than what it used to be a couple of decades ago, all because of advancement in energy and resource efficient technology and infrastructure. Since, ground water is the major source of drinking water in this region, it is important to understand how urban development affects infiltration and runoff. Using the runoff simulation model, one can select the appropriate land cover, hydrological soil type, and precipitation level to see how these three parameters affect runoff, infiltration, evaporation and precipitation.}
\end{displayquote}

Participant's reaction to these scenarios stands to reason. Their expectation was to engage with a model that provided a more personalized information on their local region's water resources, water in their taps, or streams running through their backyard. For instance, one participant wanted a model that would give him information on quality of water for drinking and recreation purposes near his residence.

\say{\textit{It's just totally personal and I was just thinking about that in terms of the water and what's going on with water management. I just want my kid to be able to fish in the stream, drink its water and swim in it. That's all I really care about, and I think that's what a lot of people care about. How we get there is stuff the smart people got to deal with but all I want to know is that...}} (Community Resident, H7).

At the same time, they viewed such a general model to be more useful for decision-makers. They believed that a lot of times, decision-makers, like local civic authorities have incorrect knowledge or information which can negatively impact the policies and decisions they make. For example, the following quote shows how a participants envisioned local water groups using such a model:

\say{\textit{This will be an integral part of what other people in the Clear Creek Watershed and Trout Welfare are trying to do, overlaying this model into Beauville Township region at least...that will definitely be helpful for them in knowing, oh this is why temperature of water at certain places along Clear Creek is rising because of the fact that because you can look at this model and know that impervious surface has gone up, they've reduced the amount of forest or grassland in the community so this should not come as a surprise to you all because the model tells you that when you do A, B is going to happen. So you just get that into local format.}} (Community Resident, H7).

\section{Discussion}
Our findings revealed social and organizational structure of different water groups, how these groups engage with the broader community, the design scenarios produced by stakeholders throughout the hackathons, technical aspects of this socio-ecological system, i.e., water data, and technical infrastructure used to manage this data. We found that members of local water groups were eager to share their data more widely, however, the tools and platforms available to them did not support those interactions. Rather than facilitating citizen water quality data practices, current tools impede and limit water groups' initiatives. Datasets owned and managed by different water groups assumed different forms stemming from varying motivations, objectives and missions of each of the described groups.  Integration of this vast and fragmented data emerged as a daunting challenge. 

Although we found that both citizen and government groups continuously take several initiatives to engage with the broader community, these practices still seem to be invisible to a lot of concerned residents who would like to get involved to help protect their watershed. For example, through our hackathons with community residents, we learned that although some participants had heard about these local water groups, they were unaware of their practices and welcomed the idea of having a platform that could help them engage with these stakeholders and participate more actively. Hence, residents are inherently motivated to get involved and learn about water data and resources. However, design concern here is for researchers to find a balance between connecting residents to water groups, and dealing with water groups' unwillingness to take a position when dealing with controversial issues.

\subsection{Design Implications}

\subsubsection{Design Outcomes of Hackathons}
The hackathons produced two important design outcomes - the water data platform mock-up, and design scenarios produced by the participants. The water data platform mock-up was initially developed as a prompt to evoke discussions in the hackathons. However, as the hackathons progressed, discussions and deliberation among participants led to the transformation of the mock-up. What started an an integrated data platform, transformed to a more comprehensive system, incorporating various functionalities. The mock-up was now a platform for community members to easily access data using different filters like, water group, water concerns, location, and time. Community members could post their stories and experiences, and also get to know about different citizen science initiatives going on in the community. The integrated data was also envisioned to be useful for informal learning. The mock-up had a page with learning tools for both adult and young members of the community. 

The benefits of integrating citizen science data has been previously recognized as an important way to address varied citizen science objectives \cite{newman2011art}. Advancement in mobile technology and increased general accessibility has facilitated volunteer participation and efficient data collection \cite{paulos2009citizen}. Smartphones now have sensors to collect varied data, the cameras with high megapixel resolution allow to capture pictures, instantaneously and seamlessly share data with scientists, and create a network of citizen scientists. For instance, Graham et al.\cite{graham2011using} developed a mobile application with various integrated tools to simplify data collection, and increase volunteer retention and engagement through online social interactions. 

Hence, the empirical investigation lead to development of a platform prototype, that was envisaged by researchers a decade ago. The mock-up morphed into a networked community-based monitoring platform with integrated data and tools. The aim of this system was to not just store data but also provide a space for both water stakeholders and community members to use this data to make decisions and learn about local water system. Successful implementation of such a mock-up has the potential to augment this community of practice. It can motivate community members to become watershed stewards and join local citizen science initiatives, and motivate school kids and kindle their interest in science and local community.

The design scenarios were a culmination of concerns of both water stakeholders and community members. The scenarios engendered by the hackathons helped identify some of the main watershed concerns, and appropriate design direction to address these concerns.  The five scenarios presented were an aggregation of all major concerns expressed by the hackathon participants. These scenarios are neither direct description of concerns raised by participants, nor a set of requirements for design. Rather, they represent a narrative description of a factual situation supported by technology \cite{nardi1992use}. The concerns of participants transformed into scenarios for design after discussion and analysis between water stakeholders and our research team. These discussions led to the conclusion that the scenarios could be best depicted through interactive simulation models. Hence, scenarios provided a concrete use of a future technology.   

In the final hackathon, we presented an interactive simulation model depicting runoff in a watershed to get feedback on usefulness of such a design from community members. We developed a scenario on how participants could use the model to understand impact of land development on groundwater. The scenario helped in providing a socio-technical walkthrough \cite{herrmann2009systems} of the design which gave us valuable feedback on feasibility of such a design. The community members had a positive reaction towards the idea of an interactive simulation model, but wanted the model to address more micro-level water concerns, such as water quality for fishing, and swimming in local streams and lakes. Hence, the scenario-based design approach helped us reflect on our initial design direction, and gave us more detailed perspective to further improve our work \cite{carrol1999five}. 

Scenario-based design is one of the most popular interactive design approach. It provides both concrete and flexible design opportunities, and helps designers cope with the dynamic changes in user requirements
Such data-driven models \cite{rosson2009scenario}. Hence, data-driven models driven by scenarios have the potential to educate the community about their local water system, help take evidence-based decisions, and encourage a more rational discourse ultimately strengthening the social capital of the community.

\subsubsection{Implications of Design Hackathons}

The design hackathons were drivers of this research. The persistent use of design hackathons as a method helped in engaging different stakeholders throughout the research and design process. Hackathons are traditionally organized with participants with technical expertise, to innovate and build systems and prototypes in a short defined period of time. These traditional hackathons are popular in the information technology industry as they provide a quick way of coming up with solutions for various technical problems \cite{nolte2018you}. The design community has also started employing hackathons as issue-oriented events where the participants can either have  technical expertise or not. Often such design hackathons are about technological explorations around a social issue \cite{lodato2016issue} and lead to production beyond technical artefacts \cite{porter2017reappropriating}. 

The design hackathons in this research didn't just lead to the production of design scenarios, but also produced valuable outcomes such as creation of a unique synergy through discussions in a highly motivated and active community. For instance, participants considered the design hackathons to be events that brought all water stakeholders and residents together and facilitated exchange of information and ideas. In addition, the healthy and constructive discussion during each event led participants to critically assess and analyze scenarios. For instance, the scenario of creating a \say{report cards} for the watershed, discussed in section 4.6, was critically analyzed by all participants during that session. One participant pointed that one should be careful while framing watershed assessments in terms of \say{report cards}, as it could have unintended consequences and implications for various other factors like, a bad score in a particular region could affect the real-estate rates. Another participant pointed that such \say{report cards} style assessments would require regular updating since it would be difficult to capture real-time continuous data in such a visualization.

Another notable aspect of the hackathons were the prompts used to evoke discussion. First, the mock-up as a prompt, not only gave feedback on our initial envisioned design of a water data platform, but also helped generate new innovative scenario-based design ideas. Second, the use of stakeholder network diagram as a prompt was fairly unique. Identifying the stakeholders and how they relate to one another, is in the background of every piece of research especially the ones that use participatory methods. It is important to know who should participate and what are the stakes of different users. Third, the simulation model as a probe helped us understand the scope and effectiveness of designing such interactive simulations to represent the scenarios. 

The hackathons in this research were organized to construct an agile environment to discuss and come up with design solutions and challenges. The exploratory nature of this community research required a participatory method rather than a traditional user-study because brainstorming and discussion sessions with different stakeholders of the community led to identification of some fundamental and original design issues of water quality in the community.

\subsubsection{Role of data and collaboration in sustainability of this community of practice}

Our findings show that collaboration among different water groups was an essential instrument in the functioning of this community of practice. Both formal and informal collaborations were drivers in sustaining the water quality and health of Clear Creek Watershed. For instance, the integrated water management plan was a collaborative effort of various water groups, which aimed at integrating every group's resources and capabilities to address water related problems in the community. However, lack of appropriate technical infrastructure impedes such federated development because there isn't any infrastructure in place that could provide easy and comprehensible access to all water data, to make rational data-driven decisions. Hence, a central data management system that supports and ameliorates these collaborative practices, has the potential to augment these future sustainable endeavors. In addition, our hackathons brought the water groups together in a distinct environment, that created synergy through discussion of problems and brainstorming appropriate design solutions. 

Most groups in the Clear Creek Watershed collect water data for different purposes. Ensuring good data quality and access to data is crucial for protecting the water system and decision-making. As we have already noted, collaboration is a big part of this community. One way to ensure successful collaboration is to have a central data management system, that aggregates and stores data for easy access by any group. Such a system will foster deliberation over data issues, and improved decision-making by civic authorities. 

A central integrated water data management system can also possibly resolve the technological infrastructure problems the water groups currently face. For example, data integration can help mitigate the inconvenience caused by lack of interoperability of databases, and inefficient exchange of data. In the current system, most groups have their own outdated database systems, and exchange of data requires manual request which can be time-consuming and inefficient. In addition, design of integrated database system also needs to incorporate appropriate QA/QC provision. Lack of appropriate QA/QC measures is one of the biggest concerns of this community, as it is in the field of citizen science. Hence, such a design exploration has the potential to enhance current monitoring and management practices, which can in-turn augment social capital in the community. 

These technical infrastructure needs of citizen science groups are not complex. Water groups  require technologies that will streamline processes like data collection, quality control and assurance of data, database, data analysis and management. Hence, such requirements do not require challenging research and inventions of artefacts and can be met by either appropriating or redesigning existing technologies.

\subsection{Significance of Studying Citizen Science in the Context of Community}

Most of the community groups studied in this research, comprised of usual community residents, who are driven by intrinsic motivation to work towards protecting their local water system. Some of the hackathon participants did not have a background in water science, but were eager to learn and engage in water monitoring activities. Our discussions with the participants and analysis of the stakeholder network revealed the complex social dynamics within which these groups operate. This hyperlocal community of government and academic institutions, local residents, and non-profit organizations, have created a self-sustaining ecosystem for themselves to monitor, protect and conserve their local water system. Hence, this community of practice has also positively affected the social capital of the community by strengthening community engagement, informal learning, and trust between stakeholders. We find this complex stakeholder network to be consistent with Coleman \cite{coleman1988social}'s 1998 analysis, i.e. the multiplexity of ties in a modern community is critical to developing and facilitating flows of social capital. Hence, the way this stakeholder network operates to protect their local watershed is consistent with \textit{community-based monitoring} as studied in environmental sciences. At the same time, the findings also revealed the role and importance of data in achieving this goal. Water data is characterized as \textit{community data} because it is an essential shared-community resource that citizens of a community collect, analyze, interpret and use to protect their watershed \cite{carroll2018strengthening}. Putting this more positively, citizen-based water monitoring and data management is a natural trajectory for CSCW and community informatics \cite{newman2012future}. Citizens are the primary stakeholders in local water systems. Water is inherently local; both a local resource and a local challenge.

Studying citizen science in a community context both complements and augments CSCW literature. It complements existing literature by reproducing insights that are consistent with previous findings \cite{tsang1999replication}, and augments by investigating this citizen science activity of water monitoring at a community level.  This empirical work identified data and technical infrastructure issues that impede efforts of these citizen groups. Issues like data quality, accessibility, usability, and technical infrastructure to manage data are some of the critical issues that are confirmatory of previous findings \cite{he2016journey, wiggins2016community, snyder2017vernacular}. Additionally, this research also found previously identified social issues such as citizen engagement, motivation and learning that affect success of a citizen science project \cite{rotman2012dynamic, reeves2017crowd}. Our analysis and conceptualization of citizen-based water monitoring at a community level complements insights of CSCW community based research. The characteristics of this community, such as the network of actors and the way water monitoring activity engages them with each other and the broader community are consistent with existing findings on community network and connectedness \cite{stoll2012between}. Goals of a local community often include building social capital through increased engagement, connectedness, and learning \cite{ackerman2004communities}. These goals are easier to achieve when they are anchored in a local community problem, resource or activity, that tends to bring people together \cite{han2014enhancing, millen2002stimulating}. 

\subsection{Limitations and Future Directions}
There are two major limitations of this study. First, this study focuses on a rather more developed, resource rich, and literate hyperlocal community of citizens. Investigating design directions with groups of water stakeholders was possible because of these specific characteristics of the community. However, we recognize that it would be difficult to generalize the findings of this study to a lesser developed communities. Secondly, the design prompts used in design hackathons may have restricted the discussions and failed to give participants the space to explore and generate ideas more organically.

As a way forward, we aim to explore the design directions of the scenarios generated in these hackathon sessions, and come up with more concrete technical designs. Future research should focus on studying how such community-based practices be translated to different types of communities as well.

\section{Conclusion}
This paper explored water quality monitoring and management as a new form of community engagement. We studied a hyperlocal community of active citizens engaged in water quality monitoring and management of their local watershed. Through a series of a unique research method called `design hackathons', we acquired a comprehensive understanding of the current practices of this hyperlocal community and possible design direction to ameliorate their current practices. This research characterized water quality monitoring and management as community-based monitoring/management and water data as community data. Such a conceptual characterization helps understand how such practices can empower citizens by increasing civic engagement and active participation, and instilling data literacy and informal learning of water data and resources. A community of data-literate and cognizant citizens working together with their civic authorities, will be better equipped to preempt water related catastrophes. Hence, CBM and community data provide a great lens for HCI researchers to empower citizens and ensure that the world doesn't see another `Day Zero'.

\section*{Acknowledgments}
We appreciate the commitment and enthusiasm of our local community collaborators for participating in this study.

\bibliographystyle{unsrt}  
\bibliography{references}

\end{document}